\documentclass[aps,twocolumn,groupedaddress,showpacs]{revtex4}
\usepackage{graphicx}
\begin{document}
\bibliographystyle{apsrev}


\title{Unusual Hall Effect in Superconducting MgB$_2$ Films: \hspace{6cm}
Analogy to High-T$_c$ Cuprates}



\author{R. Jin}
\email[]{e-mail address:jinr@ornl.gov}\affiliation{Oak Ridge
National Laboratory, Oak Ridge, TN 37831}

\author{M. Paranthaman}
\affiliation{Oak Ridge National Laboratory, Oak Ridge, TN 37831}

\author{H.Y. Zhai}
\affiliation{Oak Ridge National Laboratory, Oak Ridge, TN 37831}

\author{H.M. Christen}
\affiliation{Oak Ridge National Laboratory, Oak Ridge, TN 37831}

\author{D.K. Christen}
\affiliation{Oak Ridge National Laboratory, Oak Ridge, TN 37831}

\author{D. Mandrus}
\affiliation{Oak Ridge National Laboratory, Oak Ridge, TN 37831}


\date{\today}

\begin{abstract}
We have investigated the temperature and magnetic field dependence
of the Hall coefficient of two well-characterized superconducting
MgB$_2$ films (T$_{c0}$=38.0 K) in both the normal and
superconducting states. Our results show that the normal-state
Hall coefficient R$_H$ is positive and increases with decreasing
temperature, independent of the applied magnetic field. Below
T$_c$(H), R$_H$ decreases rapidly with temperature and changes
sign before it reaches zero. The position and magnitude at which
R$_H$ shows a minimum depends on the applied field. Quantitative
analysis of our data indicates that the Hall response of MgB$_2$
behaves very similarly to that of high-T$_c$ cuprates: R$_H$
$\propto$ T and cot$\theta_H$ $\propto$ T$^2$ in the normal state,
and a sign reversal of  R$_H$ in the mixed state. This suggests
that the B-B layers in MgB$_2$, like the Cu-O planes in high-T$_c$
cuprates, play an important role in the electrical transport
properties.
\end{abstract}

\maketitle

The recent discovery of unexpectedly high temperature
superconductivity in MgB$_2$ (T$_c$ = 39 K) \cite{naga} has
stimulated a great deal of interest in its mechanism. One of the
central issues is whether MgB$_2$ is related to other well-known
superconductors or represents a new class of superconductor.
Although superconductivity was found in other borides with the
same crystal structure as MgB$_2$ \cite{cooper,leya}, the T$_c$ of
these other materials does not exceed 0.6 K. What makes the T$_c$
of MgB$_2$ almost two orders of magnitude higher? One school of
thought proposes a phonon-mediated BCS pairing mechanism. Evidence
for this view is provided by isotope effect experiments
\cite{budko}, an isotropic energy gap \cite{kara,sharoni}, NMR
studies \cite{kote,tou}, and specific heat measurements
\cite{walti,kremer}. Evidence for unconventional superconductivity
is found in the non-BCS-like temperature dependence of both
penetration depth \cite{li,pana} and microwave surface resistance
\cite{zhukov}. Theoretically, Hirsch argues that the B-B planes in
MgB$_2$ should play a role similar to the Cu-O planes in
high-T$_c$ cuprates, and predicts hole superconductivity in
MgB$_2$ \cite{hirsch}. A recent Hall study on bulk MgB$_2$ shows a
positive normal-state Hall coefficient \cite{kang}, confirming
hole conduction in this material.

The anomalous Hall response of high-T$_c$ cuprates has been
considered as one of their most remarkable and puzzling properties
\cite{anderson}. In this Letter, we report for the first time a
surprisingly similar Hall response observed in superconducting
MgB$_2$ films. The experiments show that R$_H$ is positive in the
normal state and has a strong temperature dependence that can be
described using the model Anderson proposed for high-T$_c$
cuprates \cite{anderson}. Further analysis yields a
T$^2$-dependence of the Hall angle from T$_c$ to at least 300 K.
We also find that R$_H$ changes sign as the sample is cooled below
T$_c$; the temperature and magnetic field dependence of R$_H$ in
the mixed state resembles that observed in hole-doped high-T$_c$
cuprates. We discuss the possible origin of unusual Hall behaviors
based on theoretical models for high-T$_c$ cuprates.

The MgB$_2$ films were prepared using a precursor post-processing
approach and extensively characterized as described in Ref.
\cite{para}.  For the films used in the Hall measurements, the
zero-field resistivity indicates an onset T$_c$$^{\text
{onset}}$$=38.6$ K with a transition width of $\Delta$T=0.4 K (see
Fig.\ 2b), indicating the good quality of our samples. For the
Hall measurements, we cut the samples into rectangular shape with
dimensions of 4 mm $\times$ 2 mm. The Hall and longitudinal
resistivities were measured using a standard six-point method.
Stable low-resistance contacts were achieved by heating the sample
with fresh contacts (Epotek H-20E silver epoxy) at 120 $^\circ$C
for 4 hours.  The experiments were performed in a Quantum Design
Physical Property Measurement System using a horizontal rotator
and magnetic fields up to 8 Tesla.  In order to exclude the
longitudinal contribution due to misalignment of Hall-voltage
contacts, the Hall resistivity was derived from the antisymmetric
part of the transverse resistivity under magnetic field reversal
at a given temperature, {\it i.e.},
$\rho$$_H$=[$\rho$$_H$(+H)-$\rho$$_H$(-H)]/2. Finally, the Hall
coefficient R$_H$ is evaluated from R$_H$=$\rho$$_{H}$/H.  In the
normal state, checks were made at several temperatures to ensure
that the Hall coefficient was linear in applied current and field.

In Fig.\ 1, we show the temperature dependence (5-300 K) of R$_H$
obtained on a 0.6$\mu$m thick MgB$_2$ film. The applied field was
8 T. To emphasize the variation of R$_H$ in superconducting state,
we plot the data on a semi-logarithmic scale. Note that R$_H$
reveals strong temperature dependence in both the normal and
superconducting states. Above T$_c$(8T) $\sim$ 25 K, R$_H$ is
positive and increases with decreasing temperature. Similar to
previous observations on bulk MgB$_2$ \cite{kang}, the magnitude
of the normal-state R$_H$ is essentially two orders of magnitude
smaller than that of high-T$_c$ cuprates, reflecting a much higher
carrier concentration in MgB$_2$. More striking is the behavior of
R$_H$ in the mixed state. Below T$_c$(H), R$_H$ decreases rapidly
and changes sign from positive to negative. After reaching minimum
R$_H$$^{\text {min}}$ at T$_{\text {min}}$, it increases again
until zero is approached. The general behavior is qualitatively
the same as that observed in high-T$_c$ cuprates.

\begin{figure}
\includegraphics[keepaspectratio=true, totalheight = 2.5 in, width = 2.5 in]{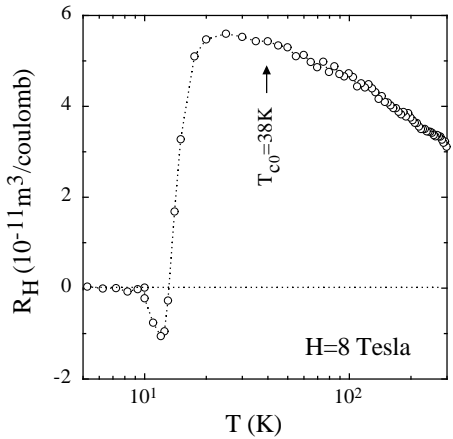}
\caption{Temperature dependence of R$_H$ at H = 8 T in a MgB$_2$
film. Note the sign change of R$_H$ in the mixed state.}
\end{figure}

We first discuss the normal-state Hall response. The temperature
dependence of R$_H$ is anomalous when analyzed in terms of
conventional transport theory. In a single-band Drude model, R$_H$
is T-independent if the anisotropy of the transport scattering
rate $\tau$$^{-1}(\bf{k})$ is independent of T, where $\bf{k}$ is
the wave vector \cite{hurt}. At present, it is unclear whether the
anisotropy in $\tau$$^{-1}(\bf{k})$ varies with temperature for
MgB$_2$. Recent band structure calculations predict that the Fermi
surface of MgB$_2$ consists of four sheets: three are hole-like
and one is electron-like \cite{kortus,shul}. With such a complex
Fermi surface, it is not surprising that R$_H$ is T-dependent.
High-T$_c$ cuprate superconductors are the examples. In the latter
system, R$_H$ is considered to be controlled by both transport
scattering rate $\tau$$^{-1}$ and transverse scattering rate
$\tau$$_H$$^{-1}$ \cite{anderson}. Within this picture, R$_H$(T)
is expected to vary as \cite{anderson}
\begin{equation}
1/R_H=aT+b,
\end{equation}
where a and b are constants. Can Eq.\ 1 also be applied to
MgB$_2$? In Fig.\ 2a, we plot the temperature dependence of the
inverse Hall coefficient 1/R$_H$ at H=8 T. Interestingly, 1/R$_H$
varies approximately linearly with T at all temperatures between
T$_c$(H) and 300 K. By fitting the data with Eq.\ 1, we obtain a =
5.06 $\times$10$^7$ C/m$^3$-K and b = 1.66 $\times$10$^{10}$
C/m$^3$. The fitting result is illustrated in Fig.\ 2a as the
solid line.

\begin{figure}
\includegraphics[keepaspectratio=true, totalheight = 4.3 in, width = 2.5 in]{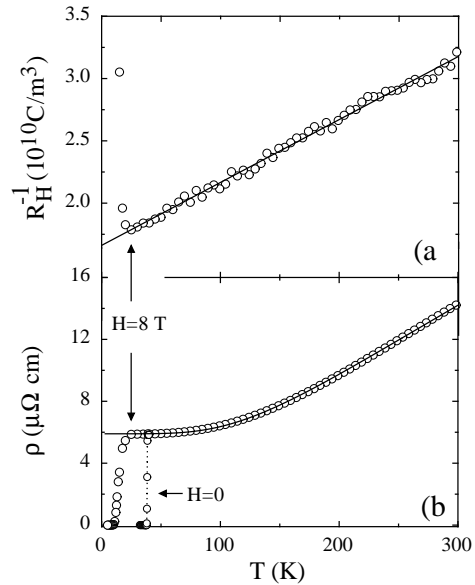}
\caption{Temperature dependence of (a) 1/R$_H$ and (b)
longitudinal resistivity $\rho$. The solid lines are fits to
experimental data using Eqs.\ 1 and 2, respectively.}
\end{figure}

The above fitting procedure demonstrates the validity of Eq.\ 1
for MgB$_2$. However, we recall that Eq.\ 1 holds if $\tau$$^{-1}$
$\propto$ T and $\tau$$_H$$^{-1}$ $\propto$ T$^2$ \cite{anderson}.
In this framework a linear T-dependence of the longitudinal
resistivity $\rho$ is expected. Fig.\ 2b shows the temperature
dependence of $\rho$ at H=0 and 8 T. Though it decreases with
decreasing T, the normal-state $\rho$ deviates from linearity
below $\sim$ 200 K and flattens as T$_c$ is approached from high
temperature. Measurements on bulk samples \cite{kang} and wires
\cite{budko} are qualitatively similar. However, we cannot exclude
the possibility that the measured $\rho$ may not represent the
in-plane resistivity $\rho$$_{ab}$. If the temperature dependence
of c-axis resistivity $\rho$$_{c}$ is distinctly different from
$\rho$$_{ab}$, the measured $\rho$ for polycrystals can deviate
significantly from $\rho$$_{ab}$. While experimental investigation
of the resistivity anisotropy is lacking, theoretical calculations
predict a nearly isotropic resistivity, which can be described by
a standard Bloch-Gr\"{u}neisen (BG) expression \cite{kong}. The BG
formula for the resistivity can be written as
\begin{equation}
\frac{\rho-\rho_0}{A}=(4\pi)^2\left(\frac{2T}{\Theta_D}\right)^5\int_0^{\Theta_D/2T}
dx \frac{x^5}{\sinh^2(x)}.
\end{equation}
where $\rho$$_0$ is the residual resistivity, A is a T-independent
constant and $\Theta_D$ is the Debye temperature. Using $\Theta_D$
= 746 K \cite{walti}, we model our resistivity data between 40 and
300 K using Eq.\ 2, yielding $\rho$$_0 $= 5.9 $\mu\Omega$ cm and A
= 0.37 $\mu\Omega$ cm. As shown in Fig.\ 2b, Eq.\ 2 (solid line)
describes our data fairly well, indicating the importance of
electron-phonon interaction in MgB$_2$. In this circumstance, it
is of surprise that the linear T-dependence of 1/R$_H$ persists
down to T$_c$(H) even in the region that $\tau$$^{-1}$ $\propto$ T
no longer holds.

Since R$_H$ depends on two scattering rates for high-T$_c$
cuprates, more attention has been paid to a simpler quantity the
Hall angle $\theta$$_H$ = tan$^{-1}$($\omega$${_c}$$\tau$$_H$)
which is independent of $\tau$, where $\omega$${_c}$ is the
cyclotron frequency. For high-T$_c$ cuprates, cot$\theta$$_H$
exhibits a quadratic temperature dependence in the normal state,
{\it i.e.},
\begin{equation}
cot\theta_H = \alpha T^2+\beta,\qquad T>T_c,
\end{equation}
where $\alpha$ and $\beta$ are constants. For MgB$_2$,
cot$\theta$$_H$=R$_H$$\cdot$H/$\rho$ is shown in Fig.\ 3 plotted
as cot$\theta$$_H$ vs. T$^2$. It should be noted that the data
fall on an approximately straight line in the whole temperature
range between T$_c$(8T) and 300 K. The solid line in Fig. 3 shows
a fit to Eq.\ 3 with $\alpha$ = 4.9e-3 K$^{-2}$ and $\beta$ = 128.
The T$^2$-dependence of cot$\theta$$_H$ and thus of
$\tau_H$$^{-1}$ suggests that $\tau$$_H$ $\not=$ $\tau$, leading
to the unconventional Hall response in MgB$_2$.

\begin{figure}
\includegraphics[keepaspectratio=true, totalheight = 2.5 in, width = 2.5 in]{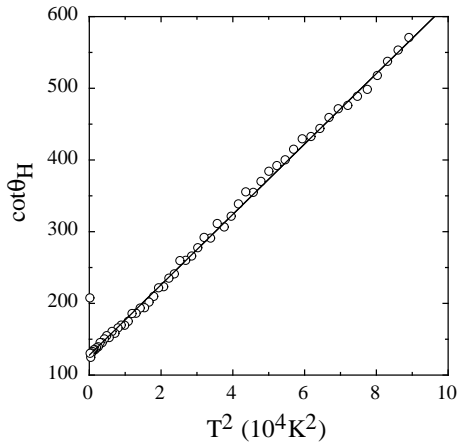}
\caption{Hall angle cot$\theta$$_H$ vs. temperature between T$_c$(H) and 300 K at H = 8 T.
 The solid line is fit to experimental data above 25 K using Eq. 3.}
\end{figure}

We now consider the Hall effect in the superconducting state.
Shown in Fig. 4a is the temperature dependence of R$_H$ at H=2, 4,
6 and 8 T, respectively. At each applied field, the R$_H$ vs. T
curve exhibits the same feature. Surprisingly, the sign reversal
of R$_H$ appears not only in high fields but also persists in low
fields. An increase of H pushes R$_H$(T) curve to lower
temperatures, {\it i.e.}, T$_{\text {min}}$ decreases with
increasing H. However, the magnitude of R$_H^{\text {min}}$ seems
to depend on H non-monotonically. It first increases and then
decreases with increasing H. Although similar features have been
observed in all high-T$_c$ cuprates and few BCS superconductors,
it is interesting that sign change of R$_H$ also occurs in
MgB$_2$. To assure that the sign change of R$_H$ is not due to
inhomogeneous superonductivity, we simultaneously measured the
longitudinal resistivity.  As shown in Fig.\ 4b, $\rho$ does not
only reveal a sharp transition in zero field but also decreases
smoothly with T in applied fields without showing any step or kink
throughout the entire transition regime. As reported previously
\cite{takano}, an increase of applied magnetic field results in a
shift of the resistive transition to lower temperatures with
appreciable broadening.

\begin{figure}
\includegraphics[keepaspectratio=true, totalheight = 4.3 in, width = 2.5 in]{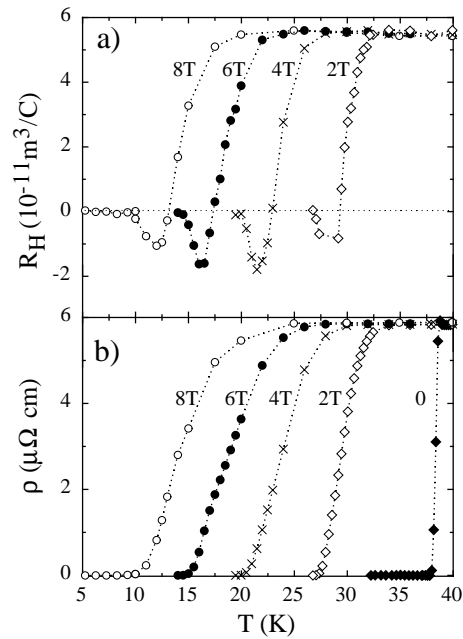}
\caption{Temperature dependence of (a) R$_H$ and (b) longitudinal resistivity $\rho$ at H = 2, 4, 6,
and 8 T.}
\end{figure}

Fig.\ 4 clearly indicates that the sign change of R$_H$ occurs
before $\rho$ reaches zero. This strongly suggests that the Hall
anomaly is a consequence of vortex dynamics. For conventional
superconductors, a phenomenological flux-flow model was developed
by Nozieres and Vinen \cite{nv} that takes into account the
hydrodynamic magnus force.  In this model, the mixed-state Hall
resistivity ($\rho_H$ = E$_y$/j$_x$) is given by:
\begin{equation}
\rho_H = (e\tau/m)\rho_n H,\qquad H<H_{c2}.
\end{equation}
Here m is the effective mass of the normal electrons, $\rho_n$ is
the normal-state longitudinal resistivity, and H$_{c2}$ is the
upper critical field. As presented in Fig.\ 5, $\rho_H$ varies
perfectly linearly with H at T = T$_{c}$(H = 0) = 38 K. Below 38
K, it departs markedly from this behavior at fields smaller than
H$_{c2}$. In this regime, $\rho_H$ increases rapidly with H after
reaching a negative peak $\rho_H$$^{\text {min}}$ at H$_{\text
{min}}$ to merge with the normal-state behavior. It is interesting
to note that the value of $\rho_H$$^{\text {min}}$ remains more or
less the same below 34 K, though H$_{\text {min}}$ increases with
decreasing temperature. It seems that a lower limit for
$\rho_H$$^{\text {min}}$ sets in at a temperature slightly below
38 K. To the best of our knowledge, such a feature has not been
seen in any other superconductors. Nevertheless, our data in Fig.\
5 demonstrates that Eq.\ 4 fails to describe the unusual
H-dependence of $\rho_H$ of MgB$_2$. In spite of many different
approaches, the mechanism responsible for the Hall sign reversal
remains controversial. For cuprate superconductors, models have
been proposed \cite{wang,ferr,harris,luo,feig} suggesting that the
sign change is related to flux pinning, backflow of thermally
excited quasiparticles, layered structure, a vortex-glass
transition, or imbalance of the electron density between the
center and the far outside the region of the vortices. In the
absence of a more detailed analysis than is available for these
scenarios, their usefulness in explaining our results remains
uncertain. However, it should be pointed out that the scaling
behavior between $\rho_H$ and $\rho$ does not seem to hold for
MgB$_2$, suggesting that the sign reversal is not a consequence of
a vortex-glass transition \cite{luo}. Given the fact that the sign
change persists at temperatures well below T$_{c0}$, the backflow
scenario should also be ruled out  \cite{ferr}. In addition, we
find that the Hall conductivity $\sigma_H$ cannot be described by
$\sigma_H$ = -C/H + DH, where C and D are positive constants.
Therefore, models based on time-dependent Ginzburg-Landau theory
\cite{dorsey,gl,kopnin} may fail to quantitatively explain our
data in mixed state. Clearly, to find out whether the Hall anomaly
arises from the layered nature of MgB$_2$, further experiments on
single crystals would be desirable.
\begin{figure}
\includegraphics[keepaspectratio=true, totalheight = 2.5 in, width = 2.5 in]{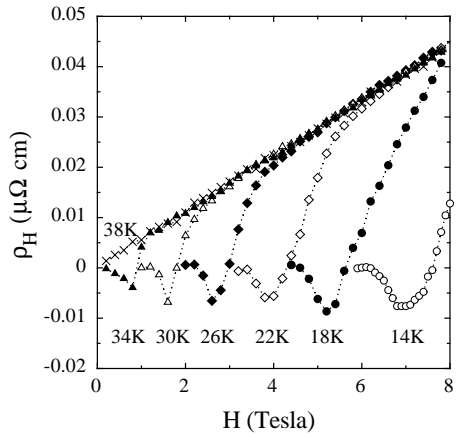}
\caption{Magnetic field dependence of the Hall resistivity $\rho$$_H$ of MgB$_2$ at T=14, 18, 22, 26,
30, 34, and 38 K.}
\end{figure}

In summary, we have measured the Hall effect on well-characterized
films of MgB$_2$ in both the normal and superconducting states.
The Hall quantities exhibit many features that are strikingly
similar to those found in high-T$_c$ cuprates. Although these
similarities may be accidental, it is more probable that these
similarities are clues to understanding the high transition
temperatures found in both MgB$_2$ and the cuprates.


\begin{acknowledgments}
We would like to thank P. Fleming for technical assistance and
E.W. Plummer and B.C. Sales for useful discussions. This work was
partly supported by the U.S. DOE, Office of Power
Technologies-Superconductivity Program, Office of EE-RE. Oak Ridge
National laboratory is managed by UT-Battelle, LLC, for the U.S.
Department of Energy under contract DE-AC05-00OR22725.
\end{acknowledgments}

\bibliography{mgb2bib.tex}

%
%

%
%

\end{document}